\documentclass{amsart}
\usepackage{graphicx}
\usepackage{color}
\usepackage[utf8]{inputenc}

\begin{document}

\title{Nonequilibrium thermodynamics: emergent and fundamental}

\author{
P. V\'an$^{1,2,3}$}

\address{$^1$Department of Theoretical Physics, Wigner Research Centre for Physics, H-1525 Budapest, Konkoly Thege Miklós u. 29-33., Hungary;\\
$^{2}$Department of Energy Engineering, Faculty of Mechanical Engineering,  Budapest University of Technology and Economics, 1111 Budapest, Műegyetem rkp. 3., Hungary\\
$^{3}$ Montavid Thermodynamic Research Group, Budapest, Hungary}



\email{van.peter@wigner.mta.hu}

\begin{abstract}{How can we derive the evolution equations of dissipative systems? What is the relation between the different approaches? How much do we understand the fundamental aspects of a second law based framework? Is there a hierarchy of dissipative and ideal theories at all?  How far can we reach with the new methods of nonequilibrium thermodynamics?}
\end{abstract}

\maketitle
\section{Non-equilibrium or nonequilibrium}

Non-equilibrium thermodynamics is a theory where the powerful methods of equilibrium are missing. Non-equilibrium thermodynamics is considered as an emergent theory; its fundamental principles, like the second law, are due to microscopic or mesoscopic properties of matter. On the other hand, nonequilibrium thermodynamics is a general framework; universal principles impose strict limits on macroscopic material properties. This way the possible micro- or mesodynamics is restricted, too. Recent developments show that the latter approach (nonequilibrium thermodynamics without the hyphen) may unify dissipative and nondissi-pative evolution. The applied principles are universal because the related methods do not require any details of the structure of the material and also in the sense that most of them are considered valid also at the more microscopic levels. Universal principles lead to universal consequences.  This topical volume introduces and overviews the efforts toward a uniform framework.

\maketitle

\section{What are the fundamental principles of dissipative evolution?}

Fundamental principles require mathematical formulations. The various approaches differ in this respect, and also the weights of the various aspects are different. The papers of the volume are instructive from this point of view. One can identify the following leading ideas:
\begin{itemize}
\item {\em Stability of thermodynamic equilibrium.} Dissipative processes tend to equilibrium; entropy is a kind of Lyapunov functional. This aspect is crucial in \cite{Ber19m,Yon19m,GrmEta19m}, important but less apparent in \cite{RomEta19m,VanKov19m}.
\item {\em Symmetric hyperbolic evolution.} A solution of evolution equations must exist when describing existing phenomena. A system of symmetric hyperbolic partial differential equations warrants that, and much more, therefore, one may consider this requirement unavoidable. This property is somehow connected to ideal systems, without dissipation, but a thermodynamic framework provides the natural variables and the necessary potential formulation. Symmetric hyperbolicity is a basic requirement in \cite{Yon19m,RomEta19m,AriEta19m} but plays an important role in \cite{OttEta19m,Jou19m}, too. 
\item {\em Spacetime embedding.} Fast and gravitating phenomena can be dissipative. Therefore, any general framework of nonequilibrium thermodynamics must have relativistic and, more importantly, covariant formulations. In nonrelativistic spacetime, this requirement, the independence of reference frames, is strongly related to material frame indifference. Dissipation does not depend on the observer. This requirement is evident in the relativistic treatment of \cite{RomEta19m}, and present also in \cite{Yon19m,VanKov19m,For19m}.
\end{itemize} 

Also, the preferred application areas influence modeling methods. For example, for {\em discrete systems}, that is homogeneous or point-like material models, the spacetime aspects are less important. These systems are treated in \cite{Ber19m,JizHan19m,Musch19m} and also partially in \cite{GrmEta19m,VanKov19m}. {\em Quantum systems} are discrete, too \cite{Ber19m,Musch19m}.

Another fundamental aspect is universality, the approach toward the emergent nature of thermodynamic principles. In nonequilibrium thermodynamics, the deviation from equilibrium can be formulated without considering the detailed structure of the material. Additional fields, frequently called internal variables, is a convenient tool  \cite{VanKov19m,For19m}. This approach proved to be very powerful in continuum theories, in particular when combined with weak nonlocality. Extended Thermodynamics introduces {\em extra fields}, too, and identifies them with dissipative fluxes \cite{Jou19m,AriEta19m,OttEta19m,GrmEta19m}.

\section{The necessary benchmarks}

Nonequilibrium thermodynamics is a metatheory, a theory of theories. More appropriately, it is a theoretical framework for describing dissipative evolution with the help of general principles. Thermodynamic ideas and concepts are applicable in various areas; they are present in very different disciplines, like quantum mechanics, continuum mechanics, electrodynamics, economics, etc. However, it is not easy to distinguish and compare the performance of the different branches. There are at least three possible benchmarks that can be applied: experimental predictivity, mathematical clarity, and backward compatibility. They help classify the different frameworks of nonequilibrium thermodynamics, like any theories of physics in general. 

\begin{figure}[!h]
\centering\includegraphics[width=5in]{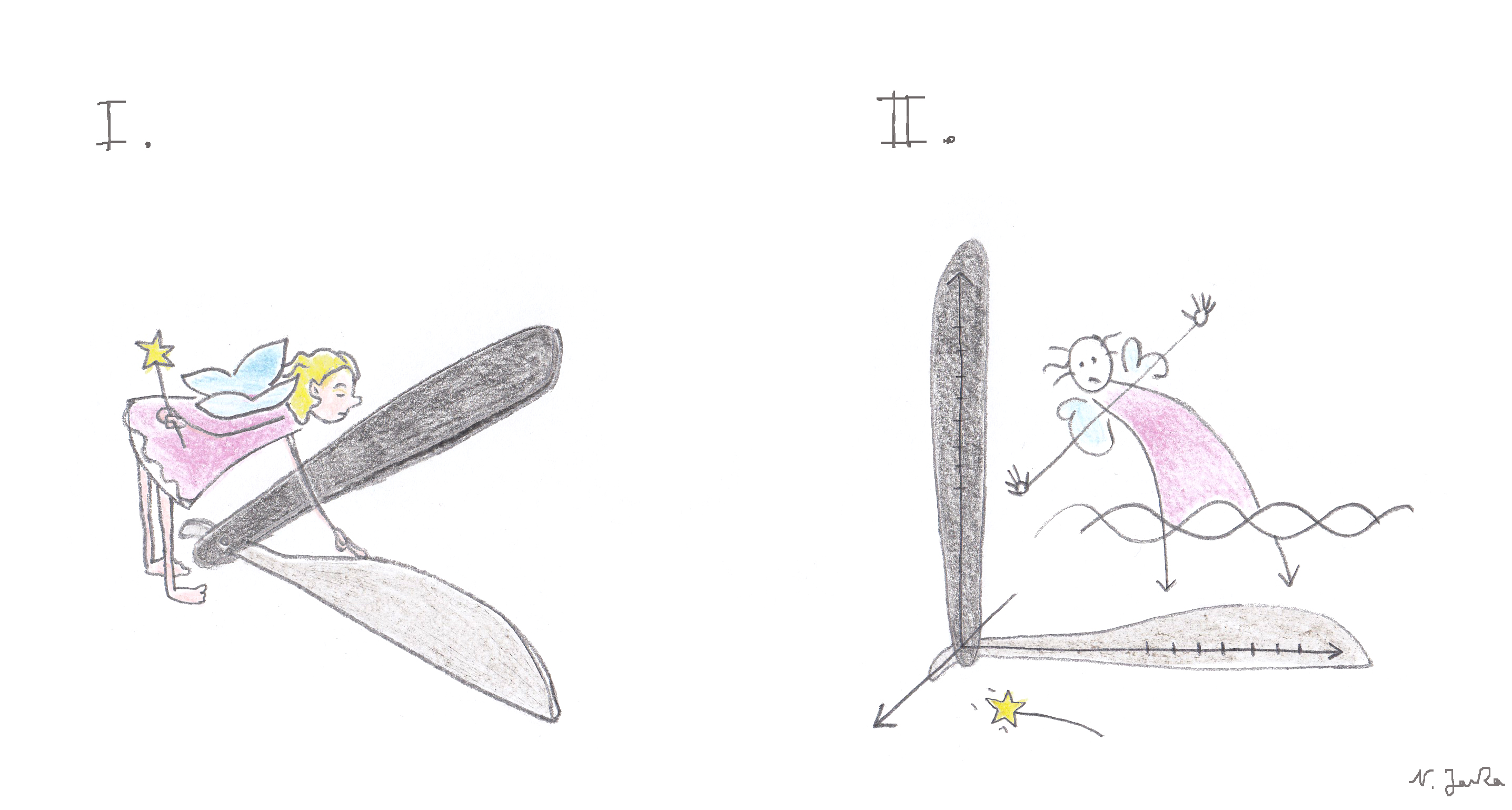}
\caption{Be careful with Ockham's razor. C: V.Gy.J.}
\label{fig_Ockham}
\end{figure}

The first, the comparison of experimental predictivity of the different branches of nonequilibrium thermodynamics does not appear in this topical issue. Also, because in this case, nonequilibrium thermodynamics is competing with other, non-thermodynamical approaches. The second, mathematical clarity is an evident heavy requirement for the approaches emphasizing symmetric hyperbolicity. It is also a general device, a magnifying glass, to identify and clean any unnecessary assumptions. However, we know well that it is hard to preserve experimental predictivity with a clean mathematical structure. Lazy intuitive concepts are useful for flexible experimental research. Ockham's razor is a dangerous device (Figure \ref{fig_Ockham}). 
The third benchmark, backward compatibility, should precede the mathematical structure. The compatibility with a {\em statistical treatment}, based on a given microscopic or mesoscopic material composition, is a clear benchmark of any general phenomenology and appears here in the papers \cite{Ber19m,JizHan19m,Abe19m}. In particular, the compatibility with the kinetic theory is important in \cite{Jou19m,AriEta19m,OttEta19m,GrmEta19m}. However, nonequilibrium thermodynamics is not necessarily emergent, from the fundamental point of view the compatibility with any nondissipative, ideal theory is expected. Ideal evolution must be a part of a general framework in the nondissipative limit. This kind of compatibility of dissipative and nondissipative frameworks is where the various branches of nonequilibrium thermodynamics most differ and where probably the best is to test the predictive power. This requirement involves the compatibility with the most fundamental and simplest equations and theories, like mechanics of point masses, gravity, and electrodynamics. Normally they are deduced from Hamiltonian variational principles. The usual approach attempts to extend the validity of {\em variational principles} to dissipative evolution \cite{YouMan99b}. In \cite{Abe19m}, we see a promising new method. On the other hand, GENERIC (General Equation for the Nonequilibrium Reversible-Irreversible Coupling) treats the {\em dissipative-nondissipative parts} on equal footing \cite{OttEta19m,GrmEta19m}. A novel approach is given in \cite{VanKov19m}, where the dissipative part is the primary, and functional derivatives, Hamiltonian dynamics is obtained from pure thermodynamic principles. 

In Table \ref{table1}. \emph{Stab., SymH., SpaceTime} denotes the stability, symmetric hyperbolicity, and spacetime arguments in the particular paper. \emph{Disc., Q, and VarPr} refer to the treatment of discrete systems, quantum thermodynamics, and the role of variational principles, respectively. \emph{EF} means internal variables or Extended Thermodynamics, that is extra fields, while \emph{Stat., IdDiss} abbreviates close relation to statistical mechanics or kinetic theory, and the aspects of dissipative-nondissipative unification. 

\vspace*{-5pt}
\begin{table}[!h]
\caption{Fundamental aspects of nonequilibrium thermodynamics, paper by paper in this topical issue (O -- represented, + -- minor representation, - -- does not appear)}
\label{table1}
\begin{tabular}{l||ccc|cccccc}
\hline
Paper & Stab. & SymH. & SpaceTime & Discr. & Q & VarPr & EF & Stat. & IdDiss\\
\hline 
Beretta         & O & - & - & O & O & - & - & + & - \\
Jizba-Hanel     & - & - & - & O & - & - & - & + & - \\
Muschik         & - & - & - & O & O & - & - & - & + \\
Abe             & - & - & - & - & - & O & - & + & - \\
Ván-Kovács      & + & - & + & + & - & O & O & - & O \\
Forest          & - & - & ? & - & - & O & O & - & - \\
Jou             & - & + & - & - & - & - & O & + & - \\
Arima et al.    & - & O & - & - & - & - & O & + & - \\
Öttinger et al. & - & + & - & - & - & - & O & + & O \\
Grmela et al.   & O & - & - & + & - & - & + & + & O \\
Romensky et al. & + & O & O & - & - & O & - & - & O \\
Wen-An Yong     & O & O & ? & - & - & - & - & - & + \\
\end{tabular}\label{tab1}
\vspace*{-4pt}
\end{table}

\section{The papers}

The papers of this topical issue are approaching the above mentioned particular benchmarks, aiming a unified understanding in their own, sometimes orthogonal looking strategies. A goal of this collection is to incite, formulate and elaborate questions that can be answered with nonequilibrium thermodynamics. I have formulated some of the possible ones below.

\begin{enumerate}
\item {\em G. P. Beretta: The fourth law of thermodynamics: steepest entropy ascent.} One of the "great laws" of nature is the second law of thermodynamics. Greatness comes from generality, according to Feynman. It is a common understanding that both equilibrium and nonequilibrium thermodynamics, and in particular the second law, is related to the stability of equilibrium. The question is exactly how? The method of steepest entropy ascent requires the existence of a metric in the state space in order to get an effective theoretical tool with a well-defined equilibrium. A metric is a strong requirement with strong consequences, applicable in stochastic and, in quantum systems. Shall we call it the fourth law of thermodynamics? 

\item {\em P. Jizba and R. Hanel: Time-Energy Uncertainty Principle for Irreversible Heat Engines.} Discrete thermodynamic bodies are usually treated independently of time. When time becomes the part of the game, e.g., we are in finite-time thermodynamics \cite{Mat05b}, then foundations are shaking, and new problems and possibilities appear. The fourth law is connected to dissipative quantum dynamics by uncertainty principles \cite{Ber19a}. However, most of the conceptual questions can be avoided, and substantial consequences can be deduced investigating Carnot like processes of the ideal gas. The analogy of thermodynamics and quantum is natural, but are these results universal?  Could it be valid beyond ideal gases and with the many various approximations?

\item {\em W. Muschik: Phenomenological Quantum Thermodynamics of Closed Bipartite Schottky Systems.} A natural approach to consider thermodynamics in a quantum framework is shown, where the weights of the density operator depend on time in the von Neumann equation. It is investigated in a discrete body-environment compound system where the second law is taken into consideration. Then the treatment leads to a natural thermodynamic foundation of Lindblad dynamics, among others. The underlying concepts are different because in nonequilibrium thermodynamics of discrete systems the concept of contact temperature helps to avoid the confusion of different concepts of equilibrium.

\item {\em S. Abe: Weak invariants in dissipative systems: Action principle and Noether charge for kinetic theory.}  Auxiliary fields are introduced in order to obtain a variational principle for the Fokker-Planck equation, where a Hamiltonian variational treatment does not work. Then it is shown that the auxiliary field is a weak invariant of the dynamics; therefore, its expectation value is conserved, while the field itself is not conserved. Then it is also shown that weak invariants are Noether charges of the particular transformations that leave the related action invariant. The proof requires Dirac's formalism of degenerate constraints.

\item {\em P. Ván and R. Kovács: Variational Principles and Thermodynamics.} Here it is shown that Hamiltonian structure and Euler-Lagrange form evolution equations do not require variational principles. Both can be easily obtained from the second law applying the classical method of separation of divergences. This way, dissipative and nondissipative evolution equations are derived together with a uniform approach. Three particular examples are treated: point masses, Cahn--Hilliard dynamics and gravity with inertia. Gravity turns out to be weakly nonlocal continuum theory. How can a weakly nonlocal theory lead to long-range interaction? 



\item {\em S. Forest: Continuum thermomechanics of nonlinear micromorphic, strain and stress gradient media.} One of the most potent approaches of continuum thermodynamics, the method of virtual power with internal variables, is applied to obtain a unified view of micromorphic, stress, and strain gradient theories in a finite deformation elastoviscoplastic framework including strain gradient plasticity. This is the first formulation of stress gradient plasticity in this framework. Relation to heat conduction is considered, too. 

The method of virtual power is a version of the exploitation of the second law with weakly nonlocal internal variables and closely related to the method of separation of divergences \cite{Mau06a}. Are these methods truly compatible?


\item {\em D. Jou: Relations between Rational Extended Thermodynamics and Extended Irreversible Thermodynamics.} Extended Thermodynamics (ET) introduces the dissipative fluxes as parts of the thermodynamic state space. This is a conceptual step regarding the relation of microscopic and mesoscopic theories. However, there are two different version of ET, the Irreversible  and the Rational one.  In this paper, a fair comparison and the explanation of the differences and their reasons are given. Is there a unified approach to keep the rigor of rational and the flexibility of irreversible extended thermodynamics?

\item {\em T. Arima, T. Ruggeri, and M. Sugiyama: Rational Extended Thermodynamics of Dense Polyatomic Gases Incorporating Molecular Rotation and Vibration.}
Rational Extended Thermodynamics is originally a theory of rarefied gases.  Motivated by the kinetic theory of polyatomic gases, a doubled hierarchy of evolution equations was introduced. Then the validity of the approach is extended for dense fluids and preserving the spacetime structure and the symmetric hyperbolicity of the equations. All what we need is a duality principle. Here a 7 field system is analysed convincingly and a particular nonequilibrium Gibbs relation is the key.

\item {\em H. C. Öttinger, H. Struchtrup, and M. Torrilhon: Formulation of moment equations for rarefied gases within two frameworks of nonequilibrium thermodynamics: RET and GENERIC.} The moment hierarchy of kinetic theory, the heart of Extended Thermodynamics (ET), has the structure of GENERIC (General Equation for the Nonequilibrium Reversible-Irreversible Coupling). However, in the latter framework, the closure of the hierarchy is problematic. Now a new idea is suggested, a particular realization of thermodynamic related Casimir symmetry. Then the compatibility of the two approaches may become complete. 

\item {\em M. Grmela, V.  Klika, and M. Pavelka: Gradient and GENERIC time evolution towards reduced dynamics.} In this work, thermodynamics is a general reduction tool between the various nondissipative dynamics. The information that is the number of degrees of freedom is decreasing between the nondissipative levels of physics \cite{Ott18b, PavEta18b}. This particular form of stability is, according to the authors, pure geometry. Several particular cases are treated. Standard pattern recognition is the part of the game, but the primary is a dynamic one. One may wonder whether there is a path toward renormalization with nonequilibrium thermodynamics?

\item {\em E. Romenski, I. Peshkov, M. Dumbser, and F. Fambri: A new continuum model for general relativistic viscous heat-conducting media.}
Symmetric Hyperbolic Thermodynamic Compatible  (SHTC) equations are based on Lagrangians. The dissipative, relaxational part is added to the ideal, reversible core. In a relativistic treatment, the challenge is to deal with the fundamental instability of the thermal interaction. The authors propose a very general framework with stability analysis and numerical examples. Is this approach compatible with Rational Extended Thermodynamics? 

\item {\em Wen-An Yong, Intrinsic Properties of Conservation-dissipation Formalism of Irreversible Thermodynamics.} The conservation-dissipation formalism (CDF) uses the Lax-Friedrich condition as a frame to get evolution equations with symmetric hyperbolic structure. This is an attractive structure, where the stability of equilibrium, the formulation of the second law, is natural and does not require a time-reversible Hamiltonian. The compatibility with other theories is considered.  


What is the exact relation between RET, CDF, SHTC equations, and GENERIC? Could we require all the conditions?
\end{enumerate}

\section{Conclusions}

A topical issue is not a consistent presentation of a field. There are at least seven named strategies in this issue: method of steepest entropy ascent (SEA), Thermodynamics with Internal Variables (TIV), Extended Irreversible Thermodynamics (EIT), Rational Extended Thermodynamics (RET), General Equation for the Nonequilibrium Reversible-Irreversible Coupling (GENERIC), Symmetric Hyperbolic and Thermodynamically Compatible (SHTC) equations, Conservation-Dissipation Formalism (CDF) of Irreversible Thermodynamics are all branches of nonequilibrium thermodynamics. Only branches are visible yet. The purpose of this topical issue is to help to understand the connections between the branches. It is not yet clear which branch is the strongest, which is the closest to the trunk, and if a trunk exists at all. The distorting perspective of our sitting position and the fresh green leaves of applications do not help. 

What are the most important among the fundamentals aspect? What is their best formulation? Is there a clarified methodology, a great blend that may lead to a uniform view of dissipative evolution? Is there a TREE behind the branches and leaves? These are the questions that push forward the various approaches of nonequilibrium thermodynamics.

\section{Acknowledgement}
The work was supported by the grants National Research, Development and Innovation Office - NKFIH 124366(124508), 123815, KH130378, TUDFO/51757/2019-ITM (Thematic Excellence Program) and FIEK-16-1-2016-0007. The research reported in this paper was supported by the Higher Education Excellence Program of the Ministry of Human Capacities in the frame of Nanotechnology research area of Budapest University of Technology and Economics (BME FIKP-NANO).

The author thank Tamás Fülöp for valuable discussions.


\bibliographystyle{unsrt}

\end{document}